\documentclass[12pt,preprint]{aastex}

\slugcomment{This manuscript is accepted by
\textbf{Ap\&SS}}

\shorttitle{Thermal instability in clumps/cores}

\shortauthors{Nejad-Asghar}

\begin{document}

\title{Thermal instability through the outer half of quasi-static spherically symmetric molecular clumps and cores}

\author{Mohsen Nejad-Asghar}

\affil{Department of Physics, University of Mazandaran, Babolsar,
Iran}

\email{nejadasghar@umz.ac.ir}

\begin{abstract}

Thermal instability (TI) is a trigger mechanism, which can explain
formation of condensations through some regions of the interstellar
clouds. Our goal here is to investigate some conditions for
occurrence of TI and formation of pre-condensations through the
outer half a quasi-static spherical molecular clump or core. The
inner half is nearly singular and ambiguous so out of scope of this
research. We consider a spherically symmetric molecular cloud in
quasi-static and thermally equilibrium state, and we use the linear
perturbation method to investigate occurrence of TI through its
outer half. The origin of perturbations are assumed to be as
Inside-Rush-Perturbation (IRP) with outward perturbed velocity at
inner region of the cloud, and Outside-Rush-Perturbation (ORP) with
inward perturbed velocity originated at the outer parts of the
cloud. The local thermal balance at the outer half of the molecular
cloud leads to a local loosely constrained power-law relation
between the pressure and density as $p \propto \rho^{1+\chi}$, where
$-0.4\lesssim \chi\lesssim 0.05$ depends on the functional form of
the net cooling function. Physically, the value of $\chi$ depends on
the power of dependence of magnetic field to the density, $\eta$,
and also on the value of magnetic field gradient, $\zeta$. For
strong magnetic field (smaller $\eta$) and/or large field gradient
(greater $\zeta$), the value of $\chi$ decreases, and vice versa.
The results show that increasing of the value of $\chi$ leads to
form a flatter density profiles at the thermally equilibrium outer
half of the molecular clump or core, and to occur more thermally
unstable IRP and ORP with smaller growth time-scales, and vice
versa.

\end{abstract}


\section{Introduction}

Molecular clouds have hierarchical structure with dense regions
nominated as clumps and cores. These dense regions are nurseries
where the stars and planets will birth. Knowing that how these dense
regions evolve to form stars and planets is therefore of crucial
importance to achieve an appropriate star formation theory.
Nowadays, with the help of infrared detectors in large arrays and on
the space telescopes, a lot of observational data are available
about the specifications of clumps and cores (e.g., Liu et al.~2019,
Brunetti \& Wilson~2019, Caselli et al.~2019, Sokol et al.~2019). We
can deduce from hierarchical structure of molecular clouds that the
clumps and cores must also have smaller condensations through their
substructures. There is some observational instances for existence
of small condensations through clumps and prestellar cores (e.g.,
Friesen et al.~2014, Kirk et al.~2017, Tokuda et al.~2018, Ohashi et
al.~2018)

There are some theoretical aspects explaining the formation of these
condensations through the molecular clumps and prestellar cores. For
example, the arm-like over-densities through clump G33.92+0.11 may
be a natural consequence of the Toomre instability, which can
fragment to form young stellar objects in shorter time-scales than
the time-scale of the global clump contraction (Liu et al.~2019).
The turbulent motions have also been proposed as being important for
dynamical evolution of star clusters during their formations from
turbulent clumps (e.g., Farias et al.~2019). Another mechanism for
the formation of over-dense regions is the effects of magnetic
fields on regulating the substructures of molecular clumps (e.g.,
Bahmani \& Nejad-Asghar~2018, Lee \& Hennebelle~2019). Since the
substructures through the molecular cores have very small masses,
and are characterized by subsonic levels of internal turbulence and
infall motions (e.g., Lee, Myers \& Tafalla~2001), the instability
processes produced via the effects of magnetic fields may be more
important than gravitational and turbulent effects (e.g., van Loo,
Falle, \& Hartquist~2007, Nejad-Asghar~2011). Thus, there may be
important to consider some non-turbulent and/or non-gravitational
instability processes such as thermal instability (TI) for formation
of small over-densities through the quiescent regions of the
molecular clumps and cores.

After the pioneer paper of Field~(1965), entitled \textit{thermal
instability}, this subject appeared to be considered as a mechanism
to explain formation of some over-densities through the interstellar
clouds (e.g., Hunter~1966, de Gouveia dal piano \& Opher~1990, Fukue
\& Kamaya~2007). Nejad-Asghar~(2011) showed that by considering the
heating due to ambipolar diffusion in the molecular clouds, the TI
criterion can be satisfied in a worthy fashion, so that the
mechanism of TI can be used to explain the formation of small
condensations in the cylindrical geometry of molecular clouds. Since
the self-gravity stratifies the gas, there must be difference
between the linear regimes of TI criterions in the different
geometries. For example, McCourt et al.~(2012) simulated the
occurrence of TI and formation of condensations through
gravitationally stratified plasmas using simplified plane-parallel
geometry, while Choudhury \& Sharma~(2016) showed that in the
non-linear simulations, there are only minor differences in cold gas
condensation for different geometries.

Since spherical geometry is more appropriate for the structure of
clumps and cores, we use the symmetric spherical geometry
approximation to investigate occurrence of TI in the linear regime.
In this geometry, the gravitational field is required to keep the
gas motionless in spite of its internal pressure, and this field
stratifies the gas into layers of varying density. If thermal
disturbances are much shorter than a scale height, one would expect
the TI to be governed by the considerations of plane-wave
approximation. Perturbations with larger scales, however, will
necessarily encompass regions of differing densities (and
temperatures). The goal of this paper is to investigate whether the
large scale perturbations through the quasi-static spherically
symmetric thermal equilibrium cloud can lead to TI process and
formation of dense regions. For this purpose, the equilibrium
profiles of the quasi-static spherically symmetric thermal
equilibrium cloud are given in \S2. In section~3, the perturbations
are applied on the basic equations with considering of net cooling
rate as a parametric power-law function. The eigenfunctions, which
obtained from perturbing cloud, are solved in \S4, and the results
are depicted. Finally, section~5 is devoted to a summary and
conclusions.

\section{Thermal equilibrium state}

In spherical polar coordinates, the usual hydrodynamic equations for
spherically symmetric molecular cloud are
\begin{equation}\label{contin}
    \frac{\partial \rho}{\partial t} + \frac{1}{r^2} \frac{\partial}{\partial
    r} \left( r^2 \rho u \right) = 0,
\end{equation}
\begin{equation}\label{moment}
    \frac{\partial u}{\partial t} + u\frac{\partial u}{\partial r}
    +\frac{1}{\rho} \frac{\partial p}{\partial r} + \frac{GM}{r^2} =0,
\end{equation}
\begin{equation}\label{mass}
    \frac{\partial M}{\partial t} + u \frac{\partial M}{\partial r} = 0,
\end{equation}
\begin{equation}\label{energy}
    \frac{\partial p}{\partial t} + u \frac{\partial p}{\partial r}
    + \gamma p \frac{1}{r^2} \frac{\partial}{\partial r} \left( r^2 u
    \right) = - \left( \gamma -1 \right) \rho \Omega,
\end{equation}
\begin{equation}\label{state}
    p = \frac{k_B}{\mu m_H} \rho T,
\end{equation}
where mass density $\rho$, the enclosed mass $M = \int_0^r 4 \pi
r^2\rho dr$, radial flow velocity $u$, thermal gas pressure $p$ and
temperature $T$ depend on the radius $r$ and time $t$; the net
cooling function is represented by $\Omega(\rho,T)$ that is
generally a complicated function of local density and temperature,
$G$ is the gravitational constant, $\gamma\approx5/3$ is the ratio
of specific heats, and $k_B$, $\mu \approx 2.3$ and $m_H$ are
Boltzmann constant, the mean molecular weight and the hydrogen mass,
respectively.

In the thermally equilibrium state, the net cooling function
$\Omega(\rho,T)$ must be zero at each radius $r$ (i.e., locally
thermal balance). To calculate the thermal balance within the
molecular clumps or cores, we need to consider heating and cooling
processes affecting the gas and the dust. Here, we use a general
parameterized form of the net cooling function as
\begin{equation}\label{netcool}
    \Omega(\rho,T) = \Lambda_0 \rho^{\xi_1} T^{\delta_1} - \Gamma_0 \rho^{\xi_2}
    T^{\delta_2},
\end{equation}
where the estimated values of the coefficients $\Lambda_0$ and
$\Gamma_0$, and the parameters $\xi_1$, $\delta_1$, $\xi_2$ and
$\delta_2$ are described in the Appendix.

We use the non-dimensional quantities $\tilde{\rho} \equiv
\rho/\rho_i$, $\tilde{p} \equiv p/p_i$, $\tilde{T} \equiv T / T_i$,
$\tilde{r} \equiv r/(\frac{p_i}{4\pi G \rho_i^2})^{\frac{1}{2}}$,
$\tilde{M} \equiv M/ 4\pi (\frac{p_i}{4\pi G
\rho_i^2})^{\frac{3}{2}} \rho_i$, $\tilde{t} \equiv t / (\frac{1}{4
\pi G \rho_i})^{\frac{1}{2}}$, $\tilde{u} \equiv u /
(\frac{p_i}{\rho_i})^{\frac{1}{2}}$, and $\tilde{\Omega} \equiv
\Omega / (\frac{4 \pi G p_i^2}{\rho_i})^{\frac{1}{2}}$, where
$\rho_i$, $T_i$, and $p_i=\frac{k_B \rho_i T_i}{\mu m_H}$ are
density, temperature, and pressure, respectively, at the outer
boundary of the molecular clump or core (i.e., intercloud medium).
In this way, the equations (\ref{contin})-(\ref{netcool}) become
\begin{equation}\label{contin2}
    \frac{\partial \tilde{\rho}}{\partial \tilde{t}} = - \frac{1}{\tilde{r}^2} \frac{\partial}{\partial
    \tilde{r}} \left( \tilde{r}^2 \tilde{\rho} \tilde{u} \right),
\end{equation}
\begin{equation}\label{moment2}
    \frac{\partial \tilde{u}}{\partial \tilde{t}}= - \tilde{u}\frac{\partial \tilde{u}}{\partial \tilde{r}}
    -\frac{1}{\tilde{\rho}} \frac{\partial \tilde{p}}{\partial \tilde{r}} -\frac{\tilde{M}}{\tilde{r}^2},
\end{equation}
\begin{equation}\label{mass2}
    \frac{\partial \tilde{M}}{\partial \tilde{t}} = - \tilde{u}\frac{\partial \tilde{M}}{\partial \tilde{r}},
\end{equation}
\begin{equation}\label{energy2}
    \frac{\partial \tilde{p}}{\partial \tilde{t}} =- \tilde{u} \frac{\partial \tilde{p}}{\partial \tilde{r}}
   - \gamma \tilde{p} \frac{1}{\tilde{r}^2} \frac{\partial}{\partial \tilde{r}} \left( \tilde{r}^2 \tilde{u}
    \right) - \left( \gamma -1 \right) \tilde{\rho} \tilde{\Omega},
\end{equation}
\begin{equation}\label{state2}
    \tilde{p} = \tilde{\rho} \tilde{T},
\end{equation}
\begin{equation}\label{netcool2}
    \tilde{\Omega}(\tilde{\rho},\tilde{T}) = \tilde{\Lambda}_0 \tilde{\rho}^{\xi_1} \tilde{T}^{\delta_1}
    - \tilde{\Gamma}_0 \tilde{\rho}^{\xi_2} \tilde{T}^{\delta_2},
\end{equation}
where $\tilde{\Lambda}_0 \equiv \Lambda_0 \rho_i^{\xi_1}
T_i^{\delta_1} / (\frac{4 \pi G p_i^2}{\rho_i})^{\frac{1}{2}}$ and
$\tilde{\Gamma}_0 \equiv\Gamma_0 \rho_i^{\xi_2} T_i^{\delta_2} /
(\frac{4 \pi G p_i^2}{\rho_i})^{\frac{1}{2}}$.

The turbulent energy sources in the molecular clouds can not be
continuously maintained (Mac Low \& Klessn~2004), i.e., the
turbulent energy will decay (Gao, Xu \& Law~2015). Assuming the
Kolmogorov scaling for eddy turbulent fluctuating velocity
(Kolmogorov~1941)
\begin{equation}\label{eddyvel}
    v_{eddy} \sim v_0 \left( \frac{l}{l_0} \right)^{1/3},
\end{equation}
where $v_0\sim 30\, \mathrm{km\,s^{-1}}$ and $l_0\sim
100\,\mathrm{pc}$ are suitable for giant molecular clouds (Gao, Xu
\& Law~2015), the turbulent decay time-scales, $l/v_{eddy}$, in the
molecular clumps (with $l\sim\, 0.3 \mathrm{pc}$) and cores (with
$l\sim\, 0.03 \mathrm{pc}$) are $7\times 10^4 \mathrm{yr}$ and
$1.5\times 10^4 \mathrm{yr}$, respectively. Since these time-scales
are comparable to the cooling time-scales (figure~2 of
Nejad-Asghar~2011), we assume that our interesting molecular clumps
and cores are approximately in the quasi static state.

The local thermal balance (i.e., $\tilde{\Omega}=0$ at each radius
$\tilde{r}$) at the outer half of the clump or core leads to a local
relation between pressure and density as $\tilde{p} = \kappa
\tilde{\rho}^{(1+\chi)}$, where $\kappa \equiv
(\frac{\tilde{\Gamma}_0}{\tilde{\Lambda}_0})^
{\frac{1}{\delta_1-\delta_2}}$ and $\chi \equiv \frac{\xi_2-\xi_1}
{\delta_1-\delta_2}$. Using this local relation between pressure and
density, the stationary ($\partial/\partial t =0$) quasi-static
($u\rightarrow 0$) state of the equations
(\ref{contin})-(\ref{state}) become
\begin{equation}\label{hydmoment}
    \frac{d\tilde{\rho}}{d\tilde{r}} = -\frac{\tilde{M}\tilde{\rho}^{(1-\chi)}}{\kappa (1+\chi) \tilde{r}^2},
\end{equation}
\begin{equation}\label{hydmass}
    \frac{d\tilde{M}}{d \tilde{r}} = \tilde{r}^2 \tilde{\rho},
\end{equation}
which can be integrated numerically (e.g., with Runge-Kutta method),
from the outer boundary of the spherical cloud $\tilde{r}=1$ with
the boundary conditions $\tilde{\rho}_{(\tilde{r}=1)} = 1$ and
$\tilde{M}_{(\tilde{r}=1)} = \tilde{M}_c$, where $\tilde{M}_c$ is
the total cloud mass in the non-dimensional scale.

According to the estimated values of the parameters $\xi_1$,
$\xi_2$, $\delta_1$, and $\delta_2$ in the Appendix, we have
$\frac{-1.9-0.3}{2.1-0.2}\approx-1.2 \lesssim \chi \lesssim
\frac{0.2-0.1}{2.1-0.2}\approx 0.05$. Since the pressure gradient
through the outer part of the clumps or cores must be a negative
value, we have the constraint $(1+\chi) \frac{d\tilde{\rho}} {d
\tilde{r}} < 0$. On the other hand, we physically expect that the
density profile of a self gravitating cloud is a decreasing function
versus the radius (i.e., $\frac{d\tilde{\rho}} {d \tilde{r}}$ must
be less than $0$), thus, we must have $\chi > -1$. Equations
(\ref{hydmoment}) and (\ref{hydmass}) show that if the value of
$\chi$ is near to $-1$, a large value of the density gradient occurs
so that the enclosed mass will be reduced to zero very rapidly. This
case cannot be physically occurred in the cloud, thus, we choose the
minimum allowed value of $\chi$ equal to $-0.5$ so that a suitable
small fraction of the total cloud mass will be enclosed in the outer
half between the inner radius $\tilde{r}_{in} \approx 0.5$ and the
outer region $\tilde{r} = 1$. Increasing of the other parameter,
$\kappa$, can only decreases the density profile, and vice versa.
Here, without loss of generality, we choose $\kappa = 1$. The
results for density profile and the enclosed mass in the outer half
of cloud, with four values of the parameter $\chi$ equal to $-0.4$,
$-0.3$, $-0.2$, and $0.05$ are depicted in the Fig.~\ref{equilib}.

\section{Perturbation analysis}

Our goal here is to investigate occurrence of thermal instability
through the outer half of a quasi-static spherically symmetric
molecular clump or core. We split each variable into unperturbed and
perturbed components; the latter is indicated by subscript '1',
while the equilibrium variables was denoted by subscript '0'. We
apply the linear perturbation analysis with time Fourier expansion,
$A_1(\tilde{r},\tilde{t}) = A_1(\tilde{r}) \exp (\omega \tilde{t})$,
on the outer half of a thermally equilibrium spherical molecular
clump or core. Time evolution in the non-linear regime is out of
scope of this paper. It is of great interest to derive the growth
rate of instability that is the real part of $\omega$. In this way,
the equations (\ref{contin2})-(\ref{state2}) can be linearized by
repeated use of the unperturbed background equations,
(\ref{hydmoment}) and (\ref{hydmass}), as follows
\begin{equation}\label{linear1}
    \omega \rho_1 = - \left( \tilde{\rho}_0' + \frac{2}{\tilde{r}} \tilde{\rho}_0\right)
    u_1 - \tilde{\rho}_0 u_1',
\end{equation}
\begin{equation}\label{linear2}
    \omega u_1 = -\frac{\tilde{M}_0}{\tilde{r}^2 \tilde{\rho}_0} \rho_1
    -\frac{1}{\tilde{r}^2} M_1 - \frac{1}{\tilde{\rho}_0} p_1',
\end{equation}
\begin{equation}\label{linear3}
    \omega M_1 = - \tilde{M}_0' u_1,
\end{equation}
\begin{equation}\label{linear4}
  \omega p_1 = - \tilde{T}_0 ( \nu_\rho - \nu_T ) \rho_1
    + \left( \frac{\tilde{\rho}_0 \tilde{M}_0}{\tilde{r}^2} - \frac{2\gamma\tilde{p}_0}{\tilde{r}} \right) u_1
    -\gamma \tilde{p}_0 u_1' - \nu_T p_1,
\end{equation}
where primes denote $\frac{d}{d\tilde{r}}$, and
\begin{equation}\label{nuT}
  \nu_T\equiv (\gamma-1)\left( \frac{\partial
\tilde{\Omega}}{\partial \tilde{T}} \right)_{\tilde{\rho}} =
(\gamma-1) \tilde{\Lambda}_0 \tilde{\rho}_0^{\xi_1}
\tilde{T}_0^{\delta_1 - 1} (\delta_1 - \delta_2),
\end{equation}
\begin{equation}\label{nuT}
  \nu_\rho \equiv (\gamma-1)\frac{\tilde{\rho}_0}{\tilde{T}_0}
\left( \frac{\partial \tilde{\Omega}}{\partial \tilde{\rho}}
\right)_{\tilde{T}} = (\gamma-1) \tilde{\Lambda}_0
\tilde{\rho}_0^{\xi_1} \tilde{T}_0^{\delta_1 - 1} (\xi_1 - \xi_2),
\end{equation}
are angular frequencies of sound waves with isochoric and isothermal
perturbations, respectively.

For small disturbances at the outer regions of the spherical cloud
(i.e., $k\tilde{r} \rightarrow \infty$, where $k$ is the
wavenumber), the plane-wave Fourier expansion, $A_1(\tilde{r}) =
A_1^d \exp (i k \tilde{r})$, with approximately homogenous medium is
suitable, as was investigated in the well-known pioneered work of
Field (1965). The Field's criterions for occurrence of linear
thermal instability with plane-wave approximation in the homogeneous
medium are
\begin{equation}\label{field}
\cases{
       a)\quad\nu_T \leqslant 0 &  for all $\nu_\rho$,\cr
       b)\quad\nu_\rho > \nu_T  & for $\nu_T\geqslant 0$, \cr
       c)\quad\nu_\rho < -(\gamma-1) \nu_T & for $\nu_T\geqslant 0$.}
\end{equation}
In our interesting region of molecular clump or core, the first
criterion cannot be occurred because we always have
$\delta_1-\delta_2 > 0$. The second and third criterions can be
represented as $\chi <-1$ and $\chi>(\gamma-1)$, respectively.
According to the mentioned values of the parameters in the Appendix,
we see that only the second criterion, $\chi<-1$, for occurrence of
linear thermal instability with plane-wave approximation in the
homogeneous medium can be reasonable (Nejad-Asghar \&
Ghanbari~2003). These classic criterions are for local linear TI in
the uniform medium. For non-uniform medium with varying density, the
classical criterions (\ref{field}) are not correct and must be
modified (Field~1965). In these cases, it may be possible for
occurrence of TI with $\chi>-1$ and/or $\chi<(\gamma-1)$, as will be
investigated further for the outer half of the spherical
clumps/cors.

For large extended (non-local) disturbances, which the sphericalness
cannot be neglected and the gravitational field stratifies the gas
into shells of varying density, we must use the general
eigenfunction method to find the thermal instability criterions. The
generic problem is to reduce the linearized equations
(\ref{linear1})-(\ref{linear4}) to a set of coupled first-order
differential equations. In this way, we have
\begin{equation}\label{diff1}
    \frac{du_1}{d\tilde{r}} = - \left( \frac{\tilde{\rho}_0'}{\tilde{\rho}_0} + \frac{2}{\tilde{r}}
    \right) u_1 -\frac{\omega}{\tilde{\rho}_0} \rho_1 ,
\end{equation}
\begin{equation}\label{diff2}
    \frac{dp_1}{d\tilde{r}}=\tilde{\rho}_0 \left( \frac{ \tilde{M}_0'}{\omega \tilde{r}^2}
    - \omega \right) u_1 -\frac{\tilde{M}_0}{\tilde{r}^2} \rho_1,
\end{equation}
\begin{equation}\label{diff3}
    \frac{d\omega}{d\tilde{r}}=0,
\end{equation}
where
\begin{equation}\label{rho1}
    \rho_1= \frac{1}{\tilde{T}_0(\nu_\rho - \nu_T - \gamma \omega)}
    \left\{ \left[ \frac{\tilde{\rho}_0 \tilde{M}_0}{\tilde{r}^2} - \frac{2\gamma \tilde{p}_0}{\tilde{r}}
    + \gamma \tilde{T}_0\tilde{\rho}_0 \left( \frac{\tilde{\rho}_0'}{\tilde{\rho}_0} + \frac{2}{\tilde{r}} \right)
    \right]u_1 - (\omega+\nu_T) p_1 \right\}.
\end{equation}
Imposing three boundary conditions for this set of three coupled
first-order differential equations turns it into an eigenvalue
problem for $\omega$. The eigenfunctions $u_1$ and $p_1$, and
consequently $\rho_1$ and $M_1$, can also be obtained with initial
guess for $\omega$ and finding its best value to satisfy the
boundary conditions.

The physical conditions of the problem present the constraints in
the boundaries $\tilde{r}_{in}$ and $\tilde{r}_{out}$. Here, we
consider two cases for occurrence of perturbations in the outer half
of clumps and/or cores. The first is that outflows from the central
region lead to occur of perturbed gas in the inner radius boundary
$\tilde{r}_{in}$ with velocity $u_{max}$. The second case is that
the perturbations are produced from inflow of material (e.g.,
outflows from neighbor clumps or cores) at the outer radius boundary
$\tilde{r}_{out}$ with inflow velocity $-u_{max}$. In both of these
models, we assume that the perturbation of pressure at the
boundaries are zero. In this way, we consider two perturbation cases
as follows:
\begin{itemize}
  \item \textit{Inside-Rush-Perturbations} (IRP) with three two-point boundary conditions
\begin{equation}\label{irp}
    u_{1(\tilde{r}_{in}=0.5)}=+u_{max},\quad p_{1(\tilde{r}_{in}=0.5)}=0,
\quad p_{1(\tilde{r}_{out}=1)}=0.
\end{equation}
  \item \textit{Outside-Rush-Perturbations} (ORP) with three two-point boundary conditions
\begin{equation}\label{orp}
u_{1(\tilde{r}_{out}=1)}=-u_{max}, \quad p_{1(\tilde{r}_{out}=1)}=0,
\quad p_{1(\tilde{r}_{in}=0.5)}=0.
\end{equation}
\end{itemize}

\section{Results}

We must solve the set of three coupled first-order differential
equations (\ref{diff1})-(\ref{diff3}) with two-point boundary value
conditions (\ref{irp}) or (\ref{orp}). There are six parameters in
the net cooling function (\ref{netcool2}) that can specify the
occurrence of thermal instability through the outer half of a
molecular clump or core: $0.1\lesssim \xi_1 \lesssim 0.3$,
$2.1\lesssim \delta_1 \lesssim 2.6$, $-1.9\lesssim \xi_2 \lesssim
0.2$, $0\lesssim \delta_2 \lesssim 0.2$, $\tilde{\Lambda}_0$ and
$\tilde{\Gamma}_0$. In the thermally equilibrium state, these six
parameters reduced into two parameters $\kappa \equiv
(\frac{\tilde{\Gamma}_0}{\tilde{\Lambda}_0})^
{\frac{1}{\delta_1-\delta_2}}$ and $\chi \equiv \frac{\xi_2-\xi_1}
{\delta_1-\delta_2}$, and we plot some interesting profiles of them
in the Fig.~\ref{equilib}. In the perturbed state, we reached to a
two-point boundary value problem (\ref{diff1})-(\ref{diff3}) with
boundary conditions (\ref{irp}) or (\ref{orp}).

Using the shooting method (e.g., Press et al.~1992) to find the
solutions of these two-point boundary value differential equations
show that the results are not sensitive to the parameters $\xi_1$,
$\delta_1$ and $\delta_2$. Thus, we choose their median values
$\xi_1\approx 0.2$, $\delta_1\approx 2.4$ and $\delta_2\approx 0.1$
for subsequent runs. Also, $u_{max}$ is chosen equal to $0.01$. The
results show that increasing of velocity perturbation amplitude
$u_{max}$ cannot change the eigenvalue $\omega$ but can modify the
eigenfunctions with amplification of their domains (see, e.g.,
Fig~\ref{umax}). According to the parameters $\kappa$ and $\chi$, we
can obtain different plots of the eigenfunctions and their relevant
eigenvalues $\omega$. By choosing $\kappa=1$, some typical
eigenfunctions $u_1$ and $p_1$, with some values of the parameter
$\chi$, are shown in the Fig.~\ref{eigenfuncs}. Finding the
solutions revealed that the IRP and ORP cases lead to the same
results for the eigenvalue $\omega$. According to the plane-fits of
cooling and heating functions in the Appendix, the coefficients
$\Gamma_0$ and $\Lambda_0$ are estimated to be in the same order so
that the parameter $\kappa$ will be near $\sim 1$. We show the
Fig.~\ref{eigenvals} for dependence of eigenvalues $\omega$ on the
parameters $\chi$ and different values of the parameter $\kappa \sim
1$.

\section{Summary and conclusions}

\begin{figure} \epsscale{1.0} \plotone{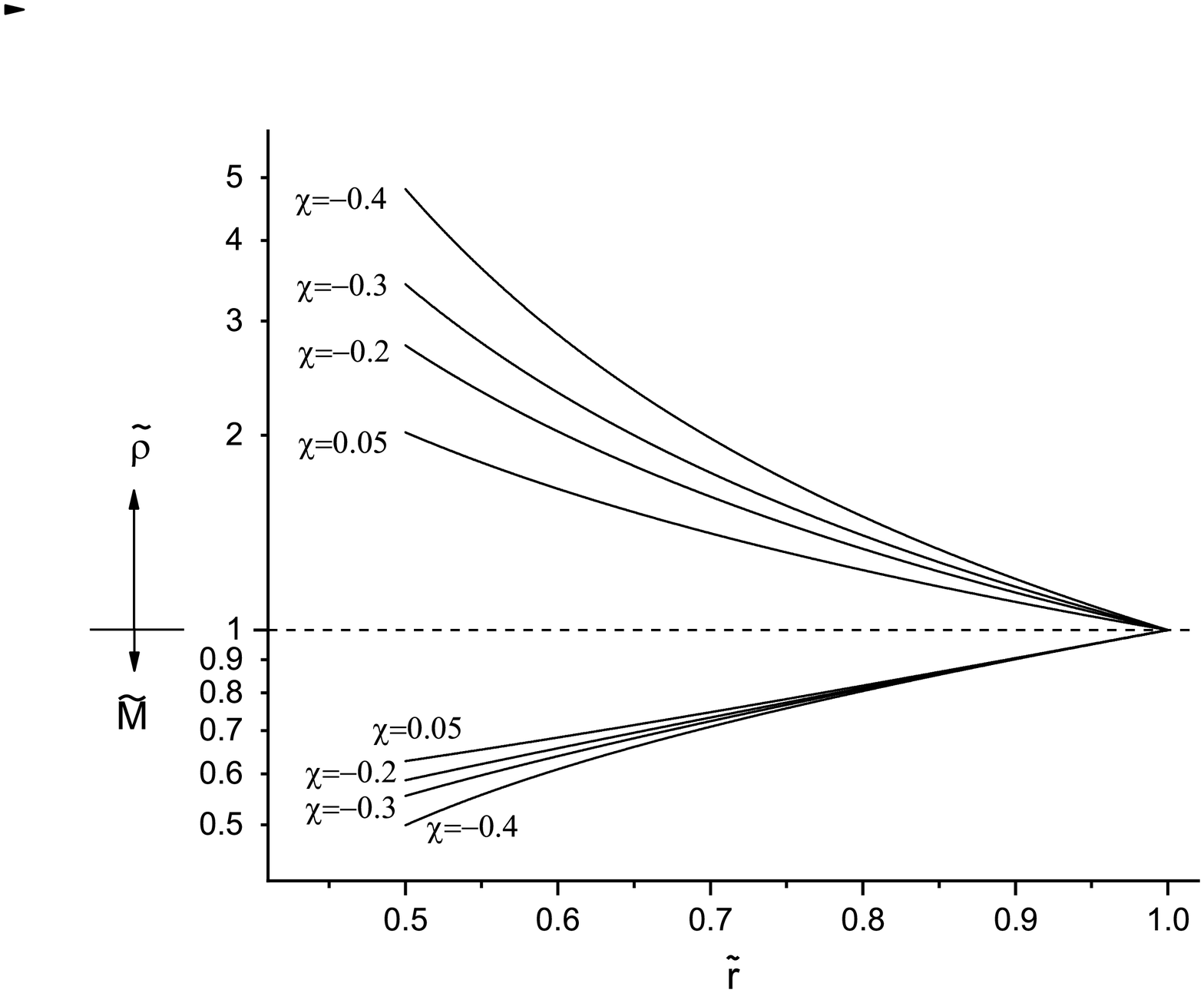}
\caption{The profiles of density and enclosed mass (in $\log_{10}$ scale) of the thermally
equilibrium states in the outer half of a quasi-static spherically
symmetric molecular clump or core. The non-dimensional quantities
are $\tilde{\rho}\equiv \rho/\rho_i$, $\tilde{r}\equiv \left(
\frac{k_B T_i}{4\pi G\mu m_H \rho_i} \right)^\frac{1}{2}$, and
$\tilde{M}\equiv 4\pi \left( \frac{k_B T_i}{4\pi G\mu m_H \rho_i}
\right)^\frac{3}{2} \rho_i$, where $\rho_i$ and $T_i$ are density
and temperature of intercloud medium, respectively.\label{equilib}}
\end{figure}

\begin{figure} \epsscale{1.0} \plotone{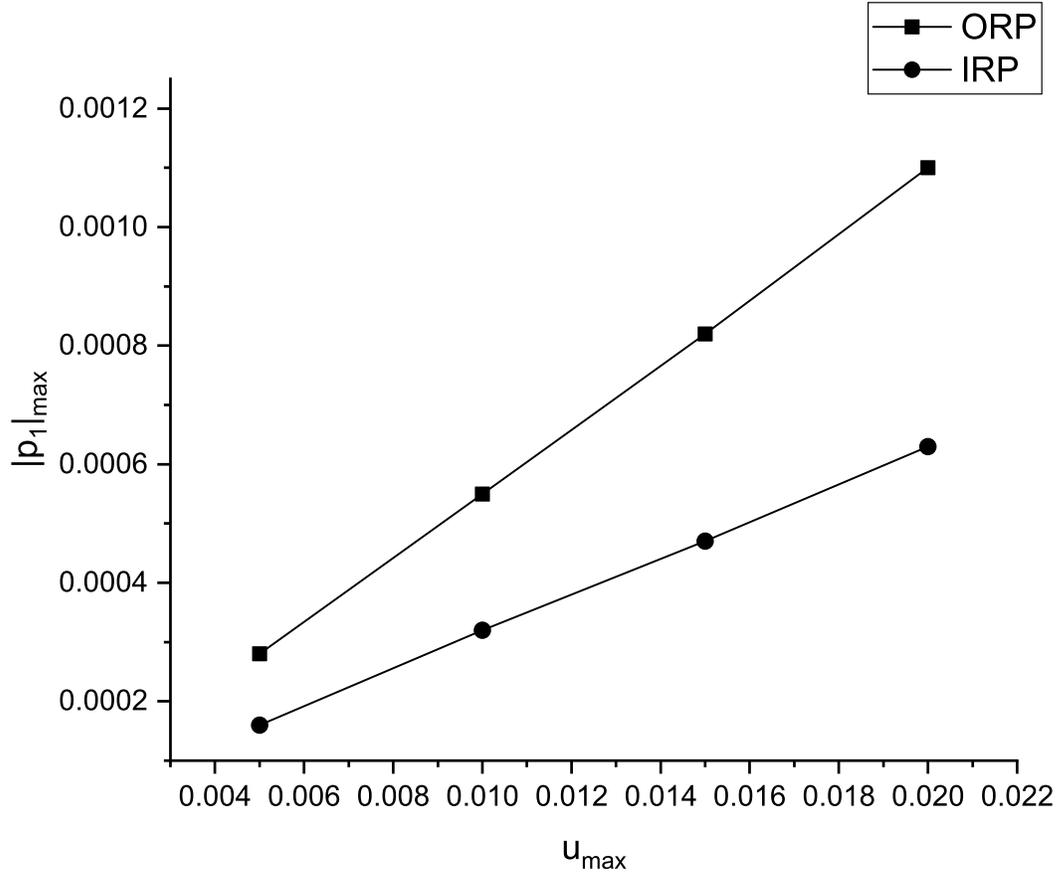}
\caption{The maximum of the absolute value of pressure perturbations
in the outer half of the clump/core with $\kappa=1$ and $\chi=-0.2$ versus
different values of the input boundary velocity $u_{max}$.\label{umax}}
\end{figure}

Molecular clouds have dense substructures, and similar over-density
substructures are also found through smaller dimensions of clumps
and cores. The existence of the small dense substructures is very
important to achieve a comprehensive theory for stars and planets
formation. In this paper, we sought to find an idea about the origin
of these small dense substructures within the molecular clumps and
cores. An idea that may seem important is the TI process. For this
purpose, we considered a molecular cloud with spherical and
quasi-static approximation and examined the conditions for thermal
instability in its outer half. The profiles of density and enclosed
mass of the thermally equilibrium states are shown in the
Fig.~\ref{equilib}.

\begin{figure}
\epsscale{.7} \center \plotone{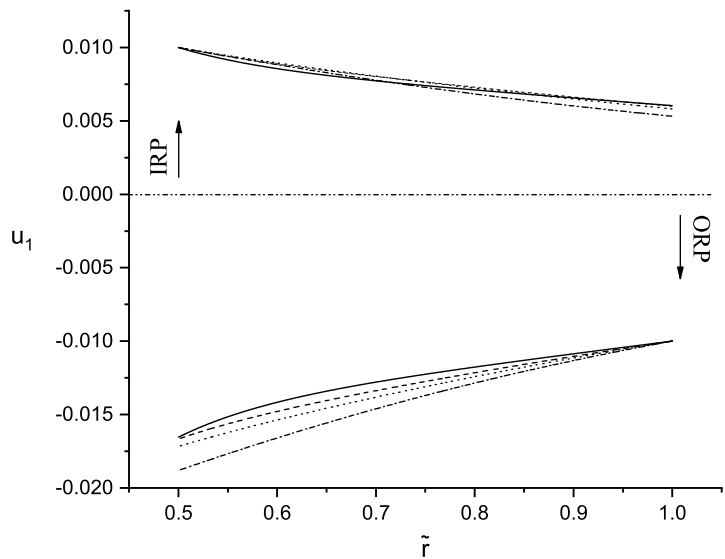}\\{(a)}\\
\epsscale{.7} \center \plotone{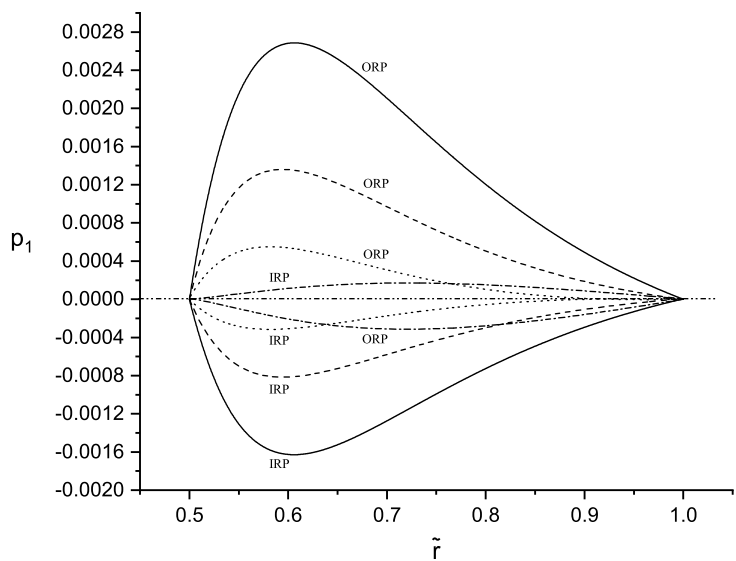}\\{(b)}\\
\caption{The eigenfunctions of (a) velocity perturbations $u_1$ and
(b) pressure perturbations $p_1$, for IRP and ORP with $\kappa=1$
and $\chi=-0.4$ (solid), $\chi=-0.3$ (dash), $\chi=-0.2$ (dot), and
$\chi=0.05$ (dash-dot). \label{eigenfuncs}}
\end{figure}

\begin{figure} \epsscale{1.0} \plotone{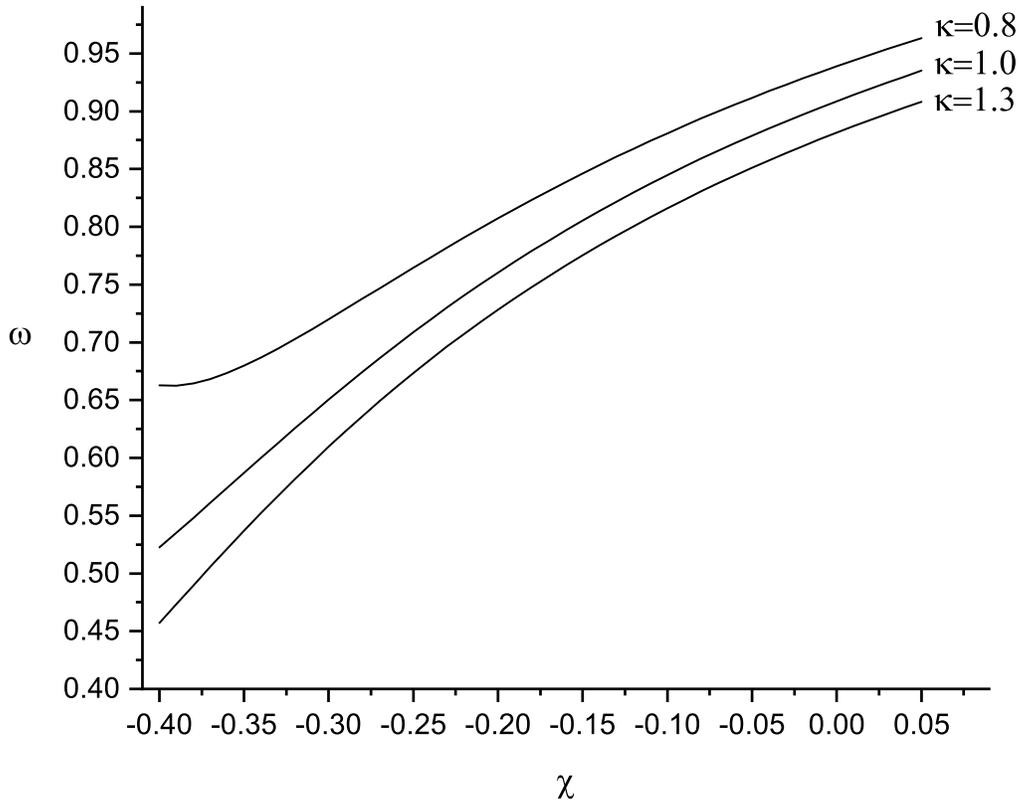}
\caption{The eigenvalues of IRP and ORP perturbations versus the
parameters $\kappa$ and $\chi$.\label{eigenvals}}
\end{figure}

To investigate the conditions for occurrence thermal instability, we
used the linear perturbation method. The most important factor in
thermal instability is the functional form of the net cooling
function. Here, we used a general parameterized power-law form for
the net cooling function (see, Appendix). We considered the origin
of the perturbations as the IRP and the ORP, and we solved the
differential equations governing the perturbations. Two important
parameters are $\chi$ and $\kappa$. Physically, the value of $\chi$
depends on the power of dependence of magnetic field to the density,
$\eta$, and the magnetic field gradient, $\zeta$. For strong
magnetic field (smaller $\eta$) and/or large field gradient (greater
$\zeta$), the minus value of $\chi$ increases, and vice versa. We
obtained the eigenfunctions and eigenvalues with the various values
of this parameter, and the results are plotted in the Figures
\ref{eigenfuncs} and \ref{eigenvals}.

It is clear from the Fig.~\ref{eigenfuncs} that the boundary
condition is satisfied in outer and inner regions (i.e.,
$p_1\rightarrow 0$ at $\tilde{r}\rightarrow 0.5\, \mathrm{and}\,
1.0$). The results show that the pressure perturbations $p_1$ are
reversed in the IRP and ORP cases. Also, absolute value of the
pressure perturbations in the inner regions are greater than the
outer regions. On the other hand, the absolute value of radial
velocity perturbations $u_1$ in the IRP cases are approximately
decreasing from inner region (place of perturbation start) to outer
area of the clump or core, while in the ORP it is an approximately
increasing function from the place of perturbation start (i.e.,
outer region). Increasing of absolute value of the velocity
perturbation from outer region to inner region is from gravitational
force acceleration. Also, the results show that increasing of $\chi$
decreases the absolute value of velocity and pressure perturbations.
We conclude from Fig.~\ref{eigenvals} that increasing of $\chi$
and/or decreasing of $\kappa$ increases the value of $\omega$ so
that the TI growth time-scales decreases. Decreasing of the
parameter $\kappa$ means the importance of heating coefficient in
relative to the cooling coefficient. Fig.~\ref{eigenvals} reveals
that considering the most valuable heating rates (specially the
heating due to ambipolar diffusion) can lead to more thermally
unstable conditions in the outer half of clumps/cores.

For better understanding of the values presented in the Figures
\ref{equilib}-\ref{eigenvals}, it is best to use the dimensional
quantities. By choosing $10\mathrm{K}$ for temperature and
$10^3\mathrm{cm^{-3}}$ for density, as references for the intercloud
medium (i.e., at the outer boundary of the spherical cloud), the
dimension of length, time, velocity, mass, and net cooling rate are
\begin{equation}\label{diml}
    0.1 \left(\frac{T_i}{10\mathrm{K}}\right)^{\frac{1}{2}}
    \left(\frac{n_i}{10^3\mathrm{cm^{-3}}}\right)^{-\frac{1}{2}}\,\mathrm{pc},
\end{equation}
\begin{equation}\label{dimt}
    5.7\times 10^5 \left(\frac{n_i}{10^3\mathrm{cm^{-3}}}\right)^{-\frac{1}{2}}\,\mathrm{yr},
\end{equation}
\begin{equation}\label{dimv}
    1.9\times 10^4 \left(\frac{T_i}{10\mathrm{K}}\right)^{\frac{1}{2}}
    \,\frac{\mathrm{cm}}{\mathrm{s}},
\end{equation}
\begin{equation}\label{dimm}
    1.0 \left(\frac{T_i}{10\mathrm{K}}\right)^{\frac{3}{2}}
    \left(\frac{n_i}{10^3\mathrm{cm^{-3}}}\right)^{-\frac{1}{2}}\,\mathrm{M_\odot},
\end{equation}
\begin{equation}\label{dimc}
    2.1\times 10^{-5} \left(\frac{T_i}{10\mathrm{K}}\right)
    \left(\frac{n_i}{10^3\mathrm{cm^{-3}}}\right)^{\frac{1}{2}}\,\mathrm{erg}\,\mathrm{g^{-1}s^{-1}},
\end{equation}
respectively.

As we can see, if we choose the temperature and density of the
intercloud medium equal to $100\mathrm{K}$ and
$10^3\mathrm{cm^{-3}}$, respectively, then we face a clump of
dimension $0.3\mathrm{pc}$ and mass $30\mathrm{M_\odot}$. On the
other hand, if we choose the temperature and density of the
intercloud medium equal to $10\mathrm{K}$ and
$10^4\mathrm{cm^{-3}}$, respectively, then we will encounter a size
and mass equal to $0.03\mathrm{pc}$ and $0.3\mathrm{M_\odot}$,
respectively, which represents a typical low-mass molecular cloud
core. According to the Fig.~\ref{eigenvals}, the growth time-scale
of TI through the outer half of these clumps and cores is
approximately in the range of $0.1-0.5\,\mathrm{Myr}$ according to
different values of the parameters $\chi$ and $\kappa$. In any case,
increasing of the parameter $\chi$ and/or decreasing of the
parameter $\kappa$ leads to occurrence of more thermally unstable
IRP and ORP with smaller growth time-scales. Therefore, as can be
seen, thermal instability can be considered as an effective
mechanism in the formation of dense regions in the outer half region
of clumps and cores.

\section*{Acknowledgments}
I appreciate the careful reading and suggested improvements by the
anonymous reviewer.

\appendix
\section{Cooling and heating rates}
Here, we briefly present the important cooling and heating processes
in the molecular clumps and cores, and the plane-fits are
approximated on the plots to determine the valuable ranges of the
parameters in the parameterized equation (\ref{netcool}). Some most
important cooling and heating mechanisms in the molecular clouds are
presented in the section~2 of Nejad-Asghar~(2011), for variety of
densities and temperatures between $10^3 < n(\mathrm{H_2}) < 10^{10}
\mathrm{cm^{-3}}$ and $10 < T < 200 \mathrm{K}$, respectively.
Nejad-Asghar~(2011) used the cooling function based on the work of
Neufeld et al.~(1995) which allowed him to include cooling from
potentially important coolants of five molecules and two atomic
species: $\mathrm{CO}$, $\mathrm{H_2}$, $\mathrm{H_2O}$,
$\mathrm{O_2}$, $\mathrm{HCl}$, $\mathrm{C}$, and $\mathrm{O}$. The
results presented in figures $3a-3d$ of Neufeld et al.~(1995) are
convenient to do rough parametrization like the equation
\begin{equation}\label{cool1}
  \Lambda_{(n,T)} = \Lambda_{(n)} \left ( \frac{T}{10 \mathrm{K}}
  \right ) ^ {\beta_{(n)}},
\end{equation}
which was introduced by Goldsmith~(2001), and the parameters are
given in the figure~1 of Nejad-Asghar~(2011).

For the heating sources in the molecular clouds, we consider heating
due to cosmic rays, turbulence heating, gravitational work, and
dissipation of magnetic energy. Nejad-Asghar~(2011) collected the
values of the heating due to cosmic rays and turbulence heating to
obtain approximately a constant value qual to
$\Gamma_{CR}+\Gamma_{TR} \sim 4.1\times 10^{-4}
\mathrm{erg\,g^{-1}\,s^{-1}}$. An estimation for the heating
produced by the self-gravitational work can be derived directly from
the rate of work per particle, $pd(n^{-1})/dt$, and is given by
\begin{eqnarray}\label{hearGR}
\nonumber  \Gamma_{GR} &\approx& \frac{p}{n t_{cn}}
\\ &\approx& 3.9\times 10^{-5} \left (\frac{T}{10
\mathrm{K}}\right ) \left ( \frac{n}{10^{6} \mathrm{cm^{-3}}} \right
)^{1/2} \quad \mathrm{erg\,g^{-1}\,s^{-1}},
\end{eqnarray}
where we have taken $dn/dt \approx n/10t_{ff}$ where $t_{ff} =
\sqrt{\frac{3 \pi}{32 G m n}}\approx 3.5 \times 10^4
\left(\frac{10^{6} \mathrm{cm^{-3}}}{n}\right)^{1/2} \mathrm{yr}$ is
the free-fall time-scale.

The dissipation of magnetic energy would be considered as another
heating mechanism, if this energy is not simply radiated away by
atoms, molecules, and grains. The major field dissipation mechanism
in the dense clouds is almost certainly ambipolar diffusion, which
was examined by Scalo~(1977) for density dependency of magnetic
field in a fragmenting molecular cloud. In the limit of low
ionization fraction (i.e., $\rho = \rho_n + \rho_i \approx \rho_n$,
where $\rho_n$ and $\rho_i$ are neutral and ion densities,
respectively), the inertia of charged particles being negligible so
that the Lorentz force $\textbf{f}_L = \frac{1}{4\pi} (\nabla\times
\textbf{B}) \times \textbf{B}$ will be balanced by the equally
important drag force per unit volume
\begin{equation}\label{dragf}
    \textbf{f}_d = \gamma_{AD} \rho_i \rho \textbf{v}_d = \gamma_{AD}
\epsilon \rho^{3/2} \textbf{v}_d,
\end{equation}
where $\gamma_{AD} \sim 3.5 \times 10^{13}
\mathrm{cm^3\,g^{-1}\,s^{-1}}$ is the collisional drag coefficient
in the molecular clouds, $\textbf{v}_d$ is the drift velocity of
ions in rest of the neutral particles, and we used the relation
$\rho_i=\epsilon\rho^{1/2}$ between ion and neutral densities in the
local ionization equilibrium state with $\epsilon\sim 3\times
10^{-16} \mathrm{g^{1/2}\,cm^{-3/2}}$ (Shu~1992). In this way, the
drift velocity of ions, with velocity $\textbf{v}_i$, relative to
the neutrals with velocity $\textbf{v}_n$, is
\begin{equation}\label{drift}
    \textbf{v}_d = \textbf{v}_i - \textbf{v}_n \approx \frac{1}{4\pi \gamma_{AD} \epsilon\rho^{3/2}}(\nabla\times\textbf{B})\times
    \textbf{B}.
\end{equation}
The magnitude of drift velocity, $v_d$, is inversely proportional to
the power of density and directly proportional to the magnetic field
strength and its gradient. Shu~(1992, Equation~27.9) used typical
values of $B \Delta B/\Delta x \sim (30 \mu \mathrm{G})^2 / 0.1
\mathrm{pc}$ to estimate the typical drift speeds through the
molecular clouds. Li, Myers \& McKee~(2012) divided the magnetic
field into a steady component and a fluctuating one, and used the
mean squared method to estimate upper limits on the drift speed.
Here, we use the parametric relation $v_d = B \zeta \rho^{-3/2}/
4\pi \gamma_{AD} \epsilon$, where the parameter $\zeta \equiv \Delta
B/ \Delta x$ is the change of magnetic field strength in the
length-scale $\Delta x$. In this way, the heating due to ambipolar
diffusion can be presented as
\begin{equation}\label{heatingAD}
  \Gamma_{AD} = \frac{\textbf{f}_d.\textbf{v}_d}{\rho} =
  \frac{B^2\zeta^2}{16\pi^2\gamma_{AD} \epsilon}
  \rho^{-2.5}.
\end{equation}
The magnetic field strength, $B$, is evaluated in the Troland \&
Crutcher~(2008) for a set of $34$ molecular cloud cores. Their
evaluations show that the magnetic field strengths are in the range
of $0.5$ to $50\,\mathrm{\mu G}$. The authors use a scaling relation
of the field strength with density, which is usually parameterized
as a power law, $B\propto \rho^\eta$ (Crutcher~2012). In the
strong-field models, $\eta \lesssim 0.5$ is predicted (e.g.,
Mouschovias \& Ciolek~1999), while for the weak magnetic fields, we
have $\eta \approx 0.66$ (Mestel~1966). Here, we use the power-law
approximation as $B\approx 100\, \mathrm{\mu G} \left(\frac{n}{10^6
\, \mathrm{cm}^{-3}}\right)^{\eta}$, where $0.3 \lesssim \eta
\lesssim 0.6$. In this way, the heating rate due to ambipolar
diffusion can be represented as
\begin{equation}\label{adheating}
    \Gamma_{AD} = 2.6 \times 10^{-8} \left( \frac{\zeta}{1\, \mathrm{\mu G} / 1\, \mathrm{mpc}} \right)^2
    \left( \frac{n}{10^{6} \mathrm{cm}^{-3}}
     \right)^{-2.5+2\eta}\, \mathrm{erg\,g^{-1}\,s^{-1}},
\end{equation}
where $\zeta \sim 1\, \mathrm{\mu G} / 1\, \mathrm{mpc}$ is suitable
for outer half of the molecular clumps and cores with strong
magnetic field and field gradient (Nejad-Asghar~2016). For weak
magnetized clumps/cores, we have $\zeta << 1\, \mathrm{\mu G} / 1\,
\mathrm{mpc}$ so that the heating due to ambipolar diffusion is
negligible.

\begin{figure}
\epsscale{1.0} \center \plotone{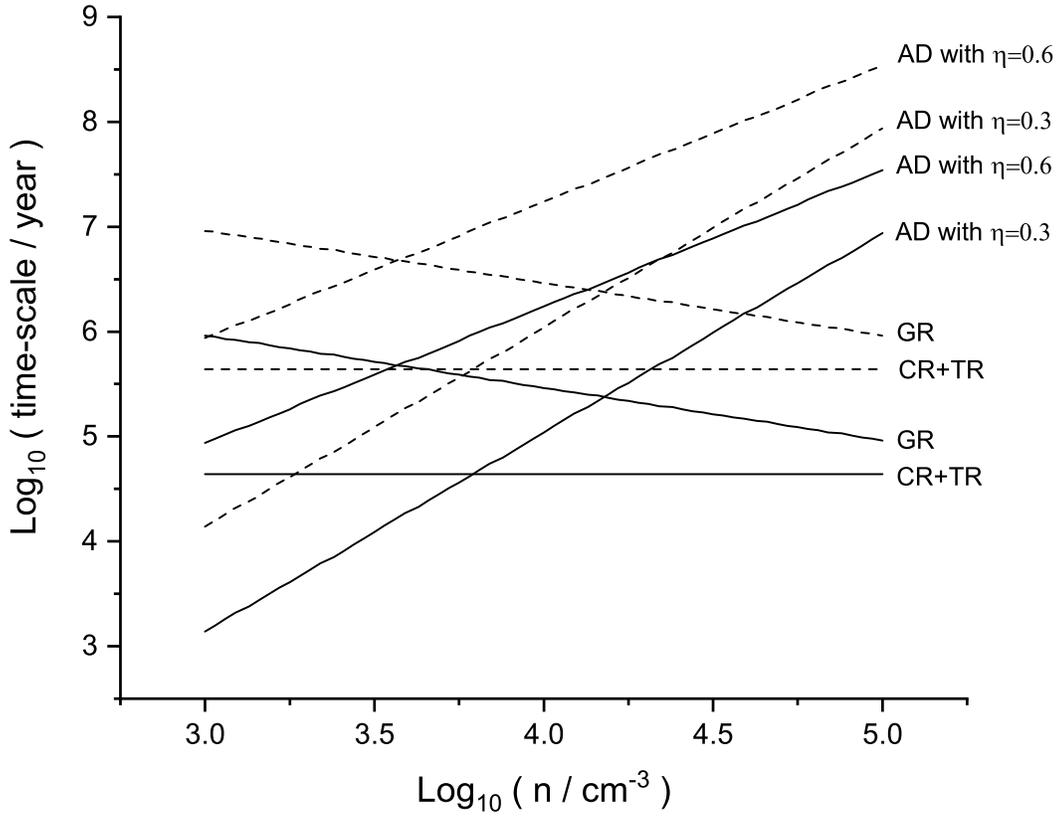} \caption{The
heating time-scales of cosmic rays and turbulence (CR+TR),
gravitational work (GR), and ambipolar diffusion (AD) with two
values of $\eta$ for $T=10\,\mathrm{K}$ (solid) and
$T=100\,\mathrm{K}$ (dash).\label{timescale}}
\end{figure}

Here, we focus our attention to the outer half of the molecular
clumps and cores with temperatures less than $200~\mathrm{K}$ and
densities between $10^3$ to $10^5\, \mathrm{cm^{-3}}$. The heating
time-scale can be approximated as $3k_BT/2\mu m_H \Gamma$ where
$\Gamma$ denotes the heating rates due to the cosmic rays and
turbulence (CR+TR), the gravitational work (GR), and the ambipolar
diffusion (AD), for evaluation of each time-scales, respectively. To
present a quantitative comparison between different heating
mechanisms, the heating time-scales are plotted in the
Fig.~\ref{timescale}. As can be seen, in the dilute regions of the
outer half of the clumps and cores with smaller densities, the
heating due to ambipolar diffusion dominates, while in the inner
regions with larger densities, importance of the ambipolar diffusion
heating is faded.

\begin{figure}
\epsscale{.52} \center \plotone{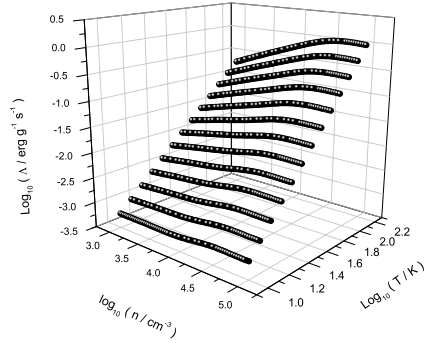}\\{(a)}\\
\epsscale{.52} \center \plotone{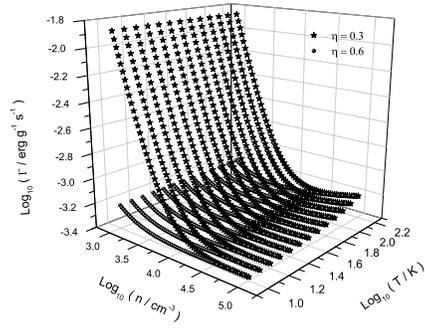}\\{(b)}\\
\epsscale{.52} \center \plotone{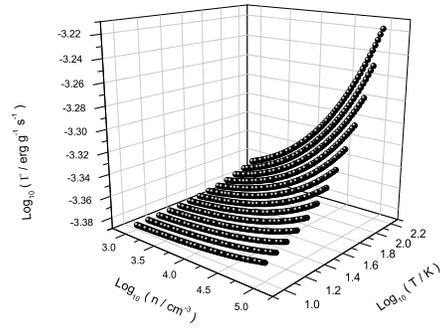}\\{(c)}\\
\caption{Three dimensional plots of the (a) cooling rate, (b)
heating rates with strong magnetic field ($\zeta \sim 1\,
\mathrm{\mu G} / 1\, \mathrm{mpc}$) and (c) heating rate with weak
magnetic field ($\zeta \rightarrow 0$).\label{coolheat}}
\end{figure}

In general, the cooling and heating rates are complicated functions
of density and temperature. The plots of cooling and heating rates,
for temperatures less than $200~\mathrm{K}$ and densities between
$10^3$ to $10^5\, \mathrm{cm^{-3}}$, are given in
Fig.~\ref{coolheat}. We can approximate some plane-fits to these
plots as follows
\begin{eqnarray}\label{planefit}
    \log \Lambda = \log \Lambda_0 + \xi_1 \log \rho + \delta_1 \log
    T,\\
    \log \Gamma = \log \Gamma_0 + \xi_2 \log \rho + \delta_2 \log
    T,
\end{eqnarray}
where $\Lambda_0$ and $\Gamma_0$ are constants in the same orders,
and the parameters $\xi_1$, $\delta_1$ and $\delta_2$ can be
estimated as $0.1\lesssim \xi_1 \lesssim 0.3$, $2.1\lesssim \delta_1
\lesssim 2.6$ and $0\lesssim \delta_2 \lesssim 0.2$. The most
important parameter is $\xi_2$ which depends on the approximated
value of $\zeta$ and the chosen value of $\eta$. Increasing the
value of $\zeta$ and/or decreasing the value of $\eta$ (i.e.,
stronger magnetic field and field gradient) leads to increase the
absolute value of slope of the fitted plane in the $\log n$
direction. For molecular clumps and cores with weak magnetic field
and field gradients, the heating rates due to cosmic rays,
turbulence and gravitational works dominate so that $\xi_2$ will be
positive. Here, we choose $-1.9\lesssim \xi_2 \lesssim 0.2$ to
consider approximately all suitable situations (in our interesting
models) of the gas in the outer half of the molecular clumps and/or
cores.


\clearpage

\clearpage

\clearpage

\clearpage

\clearpage

\clearpage

\end{document}